\documentstyle[aps,multicol,epsfig,mathrsfs,prl]{revtex}
\def\scr{\mathscr}

\def\CL{{\scr L}}

 \def\CS{{\scr S}}

\def\avg#1{\langle#1\rangle}

  \def\V0{{\bf 0}}
\def\VS{{\vec S}}  \def\VM{\vec{M}}  
 \def\Vl{{\vec l}}

\def\VO2{\vec{O}_{2k_F}}

\def\CS2{{\scr S}_{2k_F}}
\def\VCS2{\vec{{\scr S}}_{2k_F}}

\def\be{\begin{equation}}	\def\ee{\end{equation}}
\def\bea{\begin{eqnarray}}	\def\eea{\end{eqnarray}}

\begin{document}
\draft

\title{Antiferromagnetic spin ladders effectively coupled by  
one-dimensional electron liquids}
\author{W.\ Vincent Liu and Eduardo Fradkin}
\address{Department of Physics, University of Illinois at Urbana-Champaign,
1110 West Green Street, Urbana, Illinois 61801}

\date{\today}
\maketitle
\begin{abstract}
We study a model of the stripe state in strongly correlated systems
consisting of an array of antiferromagnetic spin ladders, each with
$n_{\rm{leg}}$ legs, coupled to each other through the spin-exchange
interaction to charged stripes in between each pair of ladders. The
charged stripes are assumed to be Luttinger liquids in a spin-gap
regime (Luther-Emery).  An effective interaction for a pair of
neighboring ladders is calculated by integrating out the gapped spin
degree of freedom in the charged stripe.  The low energy effective
theory of each ladder is the usual nonlinear $\sigma$-model with
additional cross couplings of neighboring ladders. These interactions
are found to favor either in-phase or anti-phase short range spin
orderings depending on whether the charge stripe is site-centered or
bond-centered as well as on its filling factor and other physical
parameters of the charged stripe.
\end{abstract}

\vskip .5cm
\vskip 0.5 cm

\begin{multicols}{2}
\narrowtext

Experiments in high $T_c$ cuprates and other doped antiferromagnets
have established that electronic stripe ordering exists in these
materials~\cite{Emery-K-T:99review}.  In this phase, the doped holes
are confined into quasi-one-dimensional structures separating locally
antiferromagnetic (AF) regions.  In addition, there is by now 
extensive evidence
suggesting that dynamical or static stripe order persist in both the normal
and superconducting states of the underdoped cuprates such as
La$_{2-x}$Sr$_x$CuO$_4$. \cite{Suzuki98:++Birgeneau99+00} These experimental
findings have led support to the picture that 
stripe order 
may be intimately related to the
mechanism of high $T_c$ superconductivity. For this, it is important
to understand the mechanisms of stripe formation.

The interplay between the charge stripe and the intervening
Mott-insulating regions has received some attention.  The spin-gap
proximity effect was found to be induced by pair hopping between the
stripe and its environment in the theory of Emery, Kivelson and
Zachar. \cite{Emery++:97} A Landau theory was also proposed by Zachar,
Kivelson, and Emery \cite{Zachar-Kivelson-Emery:98} to describe the
intermediate relation of the charge stripe order and the
incommensurate spin modulation. They argued that the stripe phase is
charge driven through continuous or first-order phase transitions:
charge order sets in first and then anti-phase spin domains set later.
Their interpretation qualitatively agrees with the experimental
observations of La$_{1.6-x}$Nd$_{0.4}$Sr$_x$CuO$_4$
\cite{LaNdSrCuO:95-97} and La$_2$NiO$_{4+\delta}$ \cite{LaNiO:93-95}.
Granath and Johannesson \cite{Granath++:99} studied a Hubbard chain
(representing the stripe) coupled to an AF spin ladder and found that
the magnetic correlations on the ladder generate a spin gap on the
stripe.  Very recently, Zachar \cite{Zachar:00pre} proposed a domain
wall (stripe) model of electron dynamics in the AF environment to
investigate the energetics of anti-phase and in-phase spin modulation.

In this paper, we attempt to find the effective interaction between
two neighboring spin regions mediated by the charge stripe. Such a
calculation is important in understanding the magnetic properties of
the (planar) doped Mott insulator in the stripe phase, and may serve
as a microscopic basis to various spin models studied recently
\cite{CastroNeto-Hone:96,Tworzydlo++:99}. Although the model used by
Zachar\cite{Zachar:00pre} differs in many ways from the picture that
we will present here, our conclusions are in full agreement with his.

It has been suggested both theoretically 
\cite{Zachar-Kivelson-Emery:98,Seibold++:98}  and experimentally
\cite{Emery-K-T:99review,LaNdSrCuO:95-97,LaNiO:93-95} 
that the charge ordering into hole rich stripes 
takes place before anti-phase
spin domain modulations set in. In this spirit we will 
consider  a model of the stripe phase of a doped AF Mott insulator in two 
dimensional
square lattice based on the following assumptions: 
1) the doped holes form an array of metallic
stripes (domain walls) lining up, say in $x$-direction; 2) between 
the charge stripes are AF insulating regions, each  
described by a
Heisenberg spin ladder with $n_{\rm{leg}}$ legs of length
$N$: 
\be
H_{S} = J\sum_j^N\sum_r^{n_{\rm{leg}}}
\vec{S}(j,r)\cdot [\VS(j+1,r) +\VS(j,r+1)]	\,.
\ee
3) the metallic stripes are Luttinger liquids 
in the Luther-Emery regime which
has a spin gap. Thus, low energy hopping process 
between the stripe and the ladders is
suppressed. \cite{Emery++:97}
Our model system qualitatively corresponds to the electronic smectic
phase of a doped Mott insulator first suggested by Kivelson, Fradkin
and Emery \cite{Kivelson-Fradkin-Emery:98}, 
but we shall not include the effects of
transverse fluctuations in the present paper.  

Now we apply the standard approach  to  a
nonlinear $\sigma$-model (see, e.g.,
Ref.~\cite{Fradkin:bk91}) for the spin ladder.
The partition function can be expressed  
as a path integral by means of the spin coherent states.
For antiferromagnetic interaction ($J>0$), the ladder is expected to be
of short-ranged N\'eel order in the ground state. Therefore, we 
split the (staggered) spin fields into a slowly
varying piece $\vec{M}$, the order parameter field, and a small
rapidly varying part $\vec{l}$:
\bea 
\VS(j,r) &\simeq& s \left[
(-)^{j+r}{\VM}(j) + a{\vec l}(j,r) \right] \,, \label{eq:StoM}
\eea
where $\VM\cdot\Vl=0$ and $\VM^2=1$. $s$ is the  spin (equal to $1/2$ for
the physical case). We have assumed that the AF
correlation length along the legs is much greater than the width of
the ladder, which allows us to treat $\VM$ constant along the rungs
\cite{Dell'Aringa++:97}.  The low energy theory of the spin
ladder  is then found to be the familiar 
nonlinear $\sigma$-model with the (Euclidean)
action \cite{Haldane:83}  
\bea
S_{0\sigma} &=& \int dx d\tau \left({1\over 2g_\sigma}
\left[ {1\over c_\sigma} (\partial_\tau\VM)^2 + c_\sigma
(\partial_x \VM)^2 \right] \right. \nonumber \\
 &&\left. + 2\pi is \sigma(n_{\rm leg})  {1\over 4\pi} \VM\cdot
(\partial_\tau\VM\times\partial_x\VM) \right)     \label{eq:L0sigma}
\eea
where the coupling constant and velocity are 
$g_\sigma = {1\over s} \left(n_{\rm{leg}} \sum_{rr^\prime}
L^{-1}_{rr^\prime} \right)^{-1/2}$, 
$c_\sigma= a J s \left(n_{\rm{leg}} /\sum_{rr^\prime}
L^{-1}_{rr^\prime} \right)^{1/2}$, 
and the matrix 
$L_{rr^\prime} \equiv 4 \delta_{rr^\prime}
+\delta_{r,r^\prime\pm1} \,.
$
The quantity $\sigma(n_{\rm leg})=0,1$ for $n_{\rm leg}$ 
even and odd respectively.
Obviously, the topological term appearing as the last in
(\ref{eq:L0sigma}) vanishes for even-leg ladders which are in a
Haldane gap phase \cite{Haldane:83}.

The charge stripe can be thought of a quasi-one dimensional electron
gas (1DEG). In the intermediate temperature regime where the system is spin
gapped but not superconducting yet, it becomes the Luther-Emery liquid
\cite{Carlson++:00} with a (Euclidean) Lagrangian in the bosonized form
\be
\CL_{\rm{1DEG}} =\CL_c+ \CL_s \,.
\ee
The charge and spin pieces are each of the sine-Gordon variety:
\bea
{\scr L}_{\alpha} &=& \frac{1}{2\pi K_\alpha}\left [ {1\over v_\alpha}
(\partial_\tau\phi_\alpha)^2 +  {v_\alpha}
(\partial_x\phi_\alpha)^2 \right] \nonumber \\ 
 && + V_{\alpha} \cos(\sqrt{8\pi} \phi_{\alpha}) \; ,  \label{eq:sineGordon}
\eea
where $\alpha = c,s$ for the charge and spin fields, respectively, and
$V_c=0$ (thus for cases of no Umklapp scattering). 
We consider only the Luther-Emery regime 
\cite{Carlson++:00} of intermediate temperature where  
$V_s$ is relevant and a spin gap, $\Delta_s$, is dynamically generated.
In this regime, the spin sine-Gordon theory is known to scale to a strong
fixed point, $K_s=1/2$, sometimes called the Luther-Emery point. In
the following, we shall restrict ourselves in performing 
calculations to this point, where some exact solution is possible. Our
analysis can be perturbatively extended to around this fixed point.

Now we are in a position to introduce the coupling between the charge
stripe and the spin ladders on both sides ($A$,$B$):
\bea
H_{\rm{int}} &=& 
{g_J \over 2}\sum_{j} [\VS_A(j,n_{leg})+\VS_B(j,1)]
 \cdot c^\dagger_\sigma(j) 
{\vec \tau}_{\sigma\sigma^\prime} c_{\sigma^\prime}(j) \, \label{eq:Hint} 
\eea
where $\vec{\tau}$ are the three Pauli matrices.
Electrons in the stripe are coupled to the
neighboring spins on the ladders at both sides. Such a spin-exchange
interaction  is quite natural
from the study of the two dimensional $t$-$J$ model for the
high $T_c$ cuprates. 
Seeking for a low energy theory, we write 
$
c_{\sigma}(j) =\sqrt{2a} \left[e^{-ik_F x} \psi_{L\sigma}(x_j) 
+ e^{+ik_F x} \psi_{R\sigma}(x_j) \right]
$,
where $\psi_{L,R}$ are slowly varying left/right moving fermion fields
with respect to the Fermi points $\mp k_F$, and 
bosonize them  \cite{Schulz++:98:1Drev}. The value of $k_F$ measures the
filling factor of the stripe.
We split each spin on the ladders
into the slowly varying order parameter field $\VM$ and the rapidly
varying field 
$\Vl$ as in (\ref{eq:StoM}).  
$H_{\rm{int}}$ splits into two separate parts, involving either $\VM$ or $\Vl$.
We shall consider the part of $\VM$ first and come back to another
afterwards. 

The part of interaction involving the order parameter fields 
amounts to an extra term to the
Lagrangian:
\bea
\CL_{\rm{int},M} &=& g_Js(-)^{n_{\rm{leg}}} \vec{\phi}_M(x)\cdot
\nonumber \\
 && \left[
   e^{-i(G_N-2k_F)x} e^{-i\sqrt{2\pi}\phi_c}   \VCS2(x) +
\rm{h.c.}  \right]    \label{eq:LM}
\eea
where 
$
\vec{\phi}_M \equiv \VM_A+\eta_s\VM_B 
$ for short-hand notation,  and 
$\VCS2(x)$ are the spin piece of the
usual $2k_F$ spin density wave order parameter (see, e.g., \cite{Voit:95}):
$
\CS2^x = {1\over \pi\alpha} \cos(\sqrt{2\pi} \theta_s)$, 
$\CS2^y = {1\over \pi\alpha} \sin(\sqrt{2\pi} \theta_s)$, and 
$\CS2^z =  {-i\over \pi\alpha} \sin(\sqrt{2\pi} \phi_s)
$ with $\alpha$ the short distance cutoff and $\theta_s$ the dual field
to $\phi_s$.
$G_N=\pi/a$ is the N\'eel ordering wave vector.
The relative phase factor, $\eta_s$,  is $+1$ ($-1$) for site-centered
(bond-centered)  stripes, arising from  staggering the spins of  two
neighbor ladders. The 
overall factor $(-)^{n_{\rm{leg}}}$ appears simply
because we have arbitrarily chosen the spin $\vec{S}_A(1,1)$ 
as the unique reference vector when staggering spins for both ladders.
As we shall see, this factor plays no physical role in the following
calculation. 

Since the spin excitations in the stripe are gapped, 
the spin degree of freedom can
be integrated out to obtain a correction term to the effective action
of $\VM$:
\be
e^{-S^\prime_{\rm{eff}}} =\int \left[
\Pi_{x,\tau} d\phi_s\right] e^{-\int dx d\tau [\CL_s +\CL_{\rm{int},M}]}
\,.
\ee
To the lowest order in $g_J$, 
\bea
S^\prime_{\rm{eff}} &=& -{\left({g_Js}\right)^2 \over 2} \int dxdx^\prime
d\tau d\tau^\prime
 \phi_M^a(x,\tau)\phi_M^b(x^\prime,\tau^\prime) \nonumber \\
&&\times
\avg{T_\tau \CS2^a(x,\tau)\CS2^b(x^\prime, \tau^\prime)}_s \nonumber \\
 && \times \Bigl\{
\Bigl( e^{-i(G_N-2k_F)x} e^{-i\sqrt{2\pi}\phi_c(x,\tau)}
+h.c.\Bigr) \nonumber\\
 &&\qquad \times  \Bigl(x, \tau \rightarrow x^\prime,\tau^\prime \Bigr) \Bigr\}
\,.
\label{eq:Seff1}
\eea
Here, the indices $a,b=x,y,z$. 
At the Luther-Emery point $K_s=1/2$, the sine-Gordon theory
(\ref{eq:sineGordon}) for the spin sector is equivalent to a theory of
free massive Dirac fermions
\cite{remark:ThirringModel}. 
The correlation function in (\ref{eq:Seff1}) can then be
calculated exactly. At zero temperature, the result is
\bea
\avg{\CS2^a(x,\tau)\CS2^b(0)}_s 
& =& {2\delta_{ab} 
\over (2\pi \xi_s)^2} \left[
K_1\left({\sqrt{x^2+v_s^2\tau^2}/\xi_s}\right)^2 \right. \nonumber \\
&& \left.  
- K_0\left({\sqrt{x^2+v_s^2\tau^2}/\xi_s}\right)^2\right]
\label{eq:SpinCo}
\eea
where $K_\nu(z)$ are Bessel functions.  Here $\xi_s$ is the
spin correlation length related to  the spin gap of the stripe,
$\xi_s=v_s/\Delta_s$, as introduced in Carlson {\it et.\ al.\/}
\cite{Carlson++:00}. (\ref{eq:SpinCo}) 
is manifestly spin rotationally invariant. 

Since the correlation function falls off at long distance ($>\xi_s$)
exponentially fast (due to the spin gap), 
we can use the operator product expansion to find an effective local
interaction. 
The product of the (normal ordered) vertex operators $e^{\pm i\sqrt{2\pi}
\phi_c}$  in 
(\ref{eq:Seff1}) at short distances yields a contribution proportional
to the identity operator of the form
\bea
&& e^{\pm i\sqrt{2\pi}\phi_c(x,\tau)} e^{\mp
i\sqrt{2\pi}\phi_c(x^\prime,\tau^\prime)}  \nonumber \\
& \simeq&
{\alpha^{K_c}\over [ (x-x^\prime)^2 + 
{v_c}^2 (\tau-\tau^\prime)^2]^{K_c\over 2}} + \cdots \label{eq:ChargeCo}
\eea
where we have dropped derivative operators.

Inserting expressions (\ref{eq:SpinCo}) and (\ref{eq:ChargeCo}) to
(\ref{eq:Seff1}), one finds $S^\prime_{\rm{eff}}$ quadratic in
$\vec{\phi}_M$ (and so $\VM$'s), in which the interaction falls off
exponentially fast. Hence, 
the long wave length limit of (\ref{eq:Seff1})
becomes (apart from an additive constant) 
\bea
S_\sigma^\prime &=& 2V\eta_s\int dx d\tau \VM_A\cdot\VM_B 
+  \cdots\mbox{(derivatives)} \label{eq:S:MAB}
\eea
The effective coupling constant $V$ is given by
\bea
V &=& -\alpha^{K_c}\left({g_Js \over 2\pi \xi_s}\right)^2
\int dx d\tau 
{\cos([G_N-2k_F]x) 
\over [ x^2 + v^2_{c}\tau^2]^{K_c\over 2}} \times
\nonumber\\ 
&& \left[
K_1\left({\sqrt{x^2+v^2_s\tau^2}\over \xi_s}\right)^2 -
 K_0\left({\sqrt{x^2+v_s^2\tau^2}\over\xi_s}
\right)^2\right] \nonumber \\
&&  \label{eq:V0}
\eea
The effective interaction term $S_\sigma^\prime$ is a correction to
the nonlinear $\sigma$-model described by the action $S_{0\sigma}$ in
(\ref{eq:L0sigma}). 
We should comment that  we have used an effective long wavelength  theory
to determine the effective coupling (\ref{eq:S:MAB}) and 
that our expressions therefore depend strongly on the short
distance cutoff.  Also, a small parameter of $G_N-2k_F$ 
cannot make our calculation more controllable for the same reason. 
This is natural
since we are computing a non-universal quantity.

So far we have left out the coupling between the
1DEG and the rapidly fluctuating fields $\Vl$, which is part of the
$H_{\rm{int}}$ in (\ref{eq:Hint}). 
Analogous to (\ref{eq:LM}), which comes from the interaction part of $\VM$, we
find an interaction term $\CL_{{\rm int},l}$ of the fields $\Vl$
coupled to both the spin density and the $2k_F$ spin density waves. 
Nevertheless, the coupling to the latter becomes negligible for the
rapidly oscillations in comparison with the former, so we ignore it. 
We then follow the same route as above 
to integrate out the spin degree of freedom in 1DEG, $\phi_s$,
and end up with effective couplings of $\Vl_A$ and $\Vl_B$.   
Adding these to the terms obtained of each spin ladder itself, we find
the total effective Lagrangian involving the fields $\Vl$ as follows 
\bea
\CL_l &=& \left[-is(\VM_A\times \partial_\tau\VM_A) \cdot
\sum_{r=1}^{n_{\rm{leg}}}\Vl_A(r) \nonumber \right. \\
&& \left. + {aJs^2\over 2} \sum_r \left( 4\Vl_A^2(r) +
2 \Vl_A(r)\cdot \Vl_A(r+1) \right) \right] \nonumber \\
 && +\Bigl[ A\rightarrow B\Bigr] \nonumber \\
 && + V^\prime a \left[ \Vl_A(n_{\rm{leg}}) +\Vl_B(1) \right]^2
\label{eq:Ll}
\eea
where the coupling constant $V^\prime =(g_J s)^2a/(4\pi v_s)$ is 
calculated exactly at the point $K_s=1/2$ 
from integrating out the spin fields $\phi_s$ of 1DEG.
Since we generally expect 
$$
{V^\prime \over J} \sim {g_J\over J} {g_J\over v_s(1/a)} \ll 1
\,,
$$
we have reduced the problem of an infinite number of weekly coupled parallel
spin ladders  to a problem of two coupled ladders in (\ref{eq:Ll}). 
\begin{figure}[htbp]
\begin{center}
\noindent
\epsfxsize=2.5in
\epsfysize=2.5in
\epsfbox{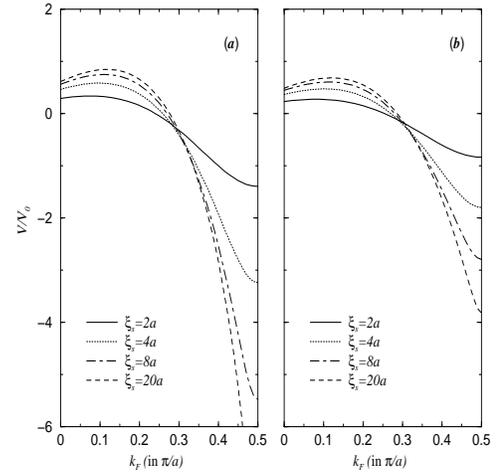}
\end{center}
\caption{
The (static) effective interaction of
$\eta_s\VM_A\cdot\VM_B$ given by the integral expression
(\ref{eq:V0}). The coupling constant $V$ is scaled with 
$V_0={s^2\over 4\pi^2}\left({g_J^2\over v_s}\right)
\left( {a^{2-K_c}\alpha^{K_c} \over \xi_s^2}\right)$.  
The lattice spacing $a$ is taken as 
the short distance cutoff such that integration in the area of $\sqrt{x^2+
v^2_s\tau^2}<a$ is excluded. Parameters are  $v_c/v_s=3$, 
and $(a)$ $K_c=1/4$; $(b)$ $K_c=3/4$.
}
\label{fig:veff0}
\end{figure}
\noindent
The effective Lagrangian $\CL_l$ is quadratic in the
fluctuating fields $\Vl$, so they can be exactly integrated out, 
leading to an effective coupling   between  the order parameter
fields of two ladders, which is of the form
\bea 
\CL_\sigma^{\prime\prime} &\sim & C^{\prime\prime} 
 (\VM_A\times\partial_\tau\VM_A)
\cdot (\VM_B\times\partial_\tau\VM_B) 
\eea 
with $C^{\prime\prime}$ a constant coefficient. Comparing with
(\ref{eq:S:MAB}), the above effective interaction is a higher order
correction to the low energy effective theory.

The static interaction of two adjacent spin ladders is solely given by
(\ref{eq:S:MAB}).  The integral in (\ref{eq:V0}) has a short distance
divergence, which will be dealt with by imposing a short distance
cutoff. A natural cutoff is the lattice spacing $a$. Thus, our results
will depend on the choice of cutoff, which is natural since this
effective coupling is controlled by local physics.  Obviously,
additional operator will yield corrections to the effective coupling
but their effects should be considerably weaker.

Fig.~\ref{fig:veff0} shows that the effective interaction $V$
changes sign when varying $k_F$, namely the filling factor of the
stripe, which for 
a free 1DEG are related by: 
$k_F= (1-\rho_{\rm h})\pi/(2a)$ with $\rho_{\rm h}$ the (linear)
density of holes in each charge stripe.

Table~\ref{tab:spin} summarizes what spin configurations are favored
by this effective interaction for different regions of $k_F$. 
Our results suggest that the effective interaction could favor
either in-phase or anti-phase spin modulation depending on whether the
stripe is site-centered or bond-centered and on its physical
parameters (such as line density). We would like to compare our
results with others in two special cases. Firstly, 
we observe from Table~\ref{tab:spin} 
that for the bond-centered stripe the in-phase spin
modulation is  favored
for small $k_F$, {\it i.e.\/}, high $\rho_{\rm h}$.
Recently, Han, Wang and Lee \cite{Han-Wang-Lee:00pre} studied the
mean-field phases of the $t$-$J$+Coulomb model numerically.
 They found that for a doping range $0.02 <x<0.14$ the
ground states are the {\it bond-centered} 
in-phase stripes, 
in which the line density of holes in each stripe
varies from $0.38$ at $x=0.095$ to $0.56$ at $x=0.14$, corresponding
to a $k_F$ from $0.31{\pi\over a}$ to $0.22{\pi\over a}$ if assuming 
a free 1DEG.  Since such a range of $k_F$ can be well considered below
some $k_F^c$, our result is consistent with that of Han,
Wang and Lee  \cite{Han-Wang-Lee:00pre} 
despite that their model and approach are  completely different from
ours. 
Secondly, we find for a
site-centered stripe configuration that a transition from anti-phase
to in-phase should occur at a critical $k^c_F$ (and thus a critical
line density). That  agrees with what Zachar \cite{Zachar:00pre} found in 
a different domain wall model, in which he estimated  the critical 
(electron) filling factor  to be $0.28<\delta_c<0.30$. 

The consistency of our result can also be checked by thinking of a
special (but unphysical) 
case: half-filled (electron) charge stripe with $k_F={\pi\over
2a}$. In this case, the $2k_F$ spin density wave of the stripe is
commensurate with the short-ranged N\'eel order (with $G_N={\pi\over
a}$) of the neighboring AF
spin ladders. Therefore, one expects that effective coupling mediated by 
the stripe should become largest in this limit. Fig.~\ref{fig:veff0} shows 
that the coupling constant $V$ is
indeed largest in magnitude for 
$k_F={\pi\over 2a}$.   
\begin{table}[htbp]
\caption{Possible spin stripe modulations favored by the effective
interaction (\ref{eq:S:MAB}). Note that $k_F^c$ denotes qualitatively
some phase transition 
point that can be read off from Fig.~\ref{fig:veff0}. } 
\begin{tabular}{l|cc}
 Stripe  
& \hspace{2em} site-centered \hspace{2em} \hss & bond-centered
\vspace{-0.5ex}\hss\\
configuration & ($\eta_s=+1$)  &  ($\eta_s=-1$) \\
\hline\hline
$0\leq k_F < k_F^c$ & anti-phase  & in-phase  \\ \hline
$k_F^c <k_F \leq {\pi\over 2a}$  & in-phase  & anti-phase  
\end{tabular}
\label{tab:spin}
\end{table}
We are grateful to Steve Kivelson for discussions and for 
pointing out ref.\ \cite{Zachar:00pre}.
During the completion of this work EF was a participant of the Program on 
High Tc Superconductivity at the Institute for Theoretical Physics, UCSB.
This work is supported in part by 
NSF grants DMR-98-17941 at UIUC and PHY94-07194 at
ITP-UCSB.

\end{multicols}

\end{document}